\documentclass[twocolumn, prl]{revtex4-1}
\usepackage{graphicx}
\usepackage{epsfig}
\begin{document}
\title{Fusing Quantum Hall States in Cold Atoms}
\author{Tin-Lun Ho}
\affiliation{Department of Physics, The Ohio State University, Columbus, OH 43210, USA}
\date{\today}

\begin{abstract}
Realizing quantum Hall states in a fast rotating Bose gas is a long sought goal in cold atom research.  
The effort is very challenging because Bose statistics fights against quantum Hall correlations.
In contrast, Fermi statistics does not cause such conflict.  
Here, we show that by sweeping the integer quantum Hall states of a spin-1/2 Fermi gas across the Feshbach resonance from the BCS side to the BEC side at a ``projection" rate similar to that in the ``projection" experiment of fermion superfluid, these states can be ``fused" into a bosonic quantum Hall states. A projection sweep means the pair association is sufficiently fast so that  the center of mass of the pair remains unchanged in the process. 
 We show that the fusion of integer fermion states with filling factor $\nu_{\uparrow}=\nu_{\downarrow}=n$ will result in a bosonic Laughlin state and Pfaffian state for $n=1$ and 2. The is due to a hidden property of the fermionic integer quantum Hall states --  for any grouping of opposite spin into pairs, their centers of mass automatically assume a bosonic quantum Hall structure. 
 \end{abstract}

\maketitle

Since the discovery of Bose-Einstein condensation in atomic gases, there have been continual efforts to realize quantum Hall (QH) states in neutral atoms.  The success of this endeavor will lead to new ways to explore quantum Hall states, such as probing particle correlations with spatial imaging or interference methods. At the same time, it will lead us to new classes of QH states inaccessible in solids, including  bosonic QH states, as well as those of high spin particles. To realize quantum Hall states, it is necessary to put  all the atoms in the lowest Landau level. This important step was accomplished for a fast rotating Bose gas in a harmonic trap by Eric Cornell's group  \cite{Cornell-LLL}. Still, reaching bosonic quantum Hall states remains a great challenge. The difficulty comes from the very nature of bosons, as Bose statistics tends to condense all particles into the same single particle level; whereas in a QH state the particles must distribute over a large number of these levels. To overcome Bose condensation, the single particle levels must be made very degenerate. Achieving such degeneracy requires the rotational frequency $\Omega$ to be so close to the frequency of $\omega_{T}$ of the harmonic trap, such that $1-\Omega/\omega_{T} \sim 1/N$, where $N$ is the number of bosons\cite{Cooper}.  This window is too narrow to achieve in current experiments unless the system contains only a few hundred bosons.

In contrast, Fermi statistics favors QH states as it naturally spreads particles into different single particle levels. Thus, for the same frequency ratio $\Omega/\omega_{T}$, it is much easier to achieve QH states with fermions than with bosons. For example, a weakly interacting spin-1/2 Fermi gas will be in an integer QH state if all the particles are in the lowest Landau level, whereas a Bose gas will remain a condensate with a vortex lattice until the aforementioned condition is met. Here, we propose a new method to create bosonic QH states.  Our scheme is to first prepare integer QH states of fermions of up and down spins on the BCS side of the Feshbach resonance, and then to sweep the system to the BEC side at the ``projection" rates such as those used in the JILA experiment to detect pair condensation\cite{Jin}. These are sweep rates too brief for significant particle motion, but fast enough to allow the formation of bosonic molecules. We show that at such sweep rate, the integer QH states up and down fermions  with filling factor $\nu_{\uparrow}=\nu_{\downarrow}=1$ and $\nu_{\uparrow}=\nu_{\downarrow}=2$ will ``fuse" into a bosonic Laughlin state and a bosonic Pfaffian state respectively. This is due to a remarkable property of integer QH states: No matter in what way the up and down spins in an integer QH state $\nu_{\uparrow}=\nu_{\downarrow}=n$ are grouped in pairs, the centers of mass of these of pairs automatically form a bosonic $n$-cluster QH state.
Thus, in a ``projection" sweep in which the motion of the centers of mass of the pairs are frozen, 
the incipient bosonic QH structures of the center of mass is revealed as the opposite spins are ``fused" into bosonic molecules at these center of mass coordinates.

{\em Cluster bosonic quantum Hall states:} We are interested in the Read-Rezayi sequence\cite{RR}, which are 
symmetrized product of $\ell$ clusters of bosonic Laughlin droplets. If all droplets have the same particle number of particles $N$, the wavefunction of the $\ell$-cluster state is 
\begin{eqnarray}
B^{(\ell)}[z] = {\cal S}  \left[  \prod_{k=0}^{\ell-1}\Phi[ z^{(k)}]^{2}  \right],
\,\,\,\,\,\, [z]= (z_1, z_2, ..z_{\ell N})  \hspace{0.3in}  \\
\Phi[w]\equiv \prod_{N\geq i>j \geq 1} (w_{i}-w _{j}), \,\,\,\,\, [w]=(w_1, w_2, ..., w_{N}), 
\hspace{0.2in}
\end{eqnarray}
where $\ell N$ is the number of bosons, $z_i=x_i+i y_i$ is the complex coordinate of the $i$-th boson, $\Phi[w]$ is the Vandermont determinant, $z^{(k)}_{i}\equiv z_{i+kN}$, $[z^{(k)}]= (z_{1+kN}, z_{2+kN}, .. , z_{(k+1)N} )$, $[z] = (z_{1}, z_{2}, ... z_{\ell N})$, and ${\cal S}$ means symmetrization with  respect to all $\ell N$ coordinates $[z]$. 
We shall omit the usual Guassian factor for simplicity. 
$B^{(1)}[z]= \Phi[z]^2$ is the bosonic Laughlin state, $B^{(2)}[z]= {\cal S} \left( \Phi[z^{(0)}]^2\Phi[z^{(1)}]^2 \right)$ is the bosonic Pfaffian state. It can be viewed as a p-wave BCS state of ``fermionized" bosons, as seen from the Cauchy identity, (proven in Appendix), 
\begin{equation}
\Phi[z] \Phi[w]= \prod_{i,j}(z_i-w_j){\rm Det}\left|\frac{1}{z_i - w_j} \right|, 
\label{Cauchy} \end{equation}
Eq.(\ref{Cauchy}) allows us to rewrite $B^{(2)}[z]$ as 
\begin{equation}
B^{(2)}[z] = {\cal S} \left( \Phi[z] \prod_{i=1}^{N}\frac{1}{z_{i}- z_{i+N}} \right). 
\label{B2} \end{equation}
The antisymmetric property of $\Phi[z]$ make the bosons in the product exchange like fermions. Consequently, the  product of pairs in Eq.(\ref{B2})is antisymmetrized, and  
becomes a fermionic p-wave BCS state. All cluster QH states with $\ell \geq 1$ are non-abelian\cite{RR}.

{\em Projection across Feshbach resonance:} In cold atoms, two fermions of opposite spin can be associated into a tightly bound pair when brought across a Feshbach resonance by varying an external magnetic field. Far from resonance on the BCS side, the interaction between fermions is weakly attractive. In high angular momentum limit, the ground states of a spin-1/2 Fermi gas are integer 
QH states. The question is what happens to these states when the magnetic field sweeps across the Feshbach resonance. 

For two fermions in a harmonic trap, the center of mass motion is unaffected by the sweep, while the motion of the relative coordinate depends on the sweep rate. For slow sweep, the wavefunction will evolve adiabatically from an extended state to a tightly bound molecule\cite{Busch}. For faster sweep, the extended states on the BCS side can still be ``projected" into small pairs on the BEC side. Experiments in the many-body case showed that ``when the projecting magnetic-field sweep is completed on a timescale that allows molecule formation but is still too brief for particles to collide or move significantly in the trap, the projection always results in 60$\%$ to 80$\%$ of the atom sample appearing as molecules"\cite{Jin}. {\em At such sweep rate, the centers of mass of the fermion pairs are forzen.} This is reflected by the fact that the quantum state of the bosonic molecules  emerging on the BEC side depends on  the phase coherence of  fermion state prior to the sweep.  
A fermion superfluid consisting of pairs of identical (zero) momentum will project onto a  Bose condensate with zero momentum, whereas a normal fermion state without pair coherence will project onto a Bose gas uncondensed\cite{Jin}. It is important to note that when passing through the resonance, the wavefunction of the relative coordinate must include many higher landau levels in order to construct a narrow bound state. In contrast, the center of mass coordinates remain in the lowest Landau level.

{\em Fusing of the integer fermion QH state $\nu_{\uparrow}=\nu_{\downarrow}=1$ under projection sweep:} We first discuss some hidden properties of integer QH states. For the integer QH state of a single component Fermi gas with filling factor $\nu=\ell$, its wavefunction $\Psi^{(\ell)}$ can be obtained by first constructing the Slater determinant $L_{k}$ of the $k$-th Landau level, $k=0,1, .., \ell-1$ and then antisymmetrizing the product of all occupied levels, $\Psi^{(\ell)}= {\cal A} (\prod_{k=0}^{\ell-1} L_{k}$). 
Noting that the single particle wavefunction of the $k$-th Landau level is $z^{\ast}{\cal P}_{m}(z)$, where ${\cal P}_{m}(z)$ is a polynomial of degree $m$, the Slater determinant for $N$ fermions in this level is  $L_{k}[z]= \Phi[z] \prod_{i=1}^{N} z_{i}^{k \ast}$. Thus, if there are $N$ fermions in all 
$\ell$ levels, the wavefunction is   
\begin{equation}
\Psi^{(\ell)}[z]= {\cal A} \left[ \prod_{k=0}^{\ell-1} \Phi[z^{(k)}] \prod_{i=1}^{N}\left( z^{(k)}_{i}\right)^{\ast k}  \right] , 
\label{ell} \end{equation}
where $[z]=(z^{}_1, z^{}_2, ..., z^{}_{\ell N} )$,  $z^{(k)}_{i} = z^{}_{i+k N}$, 
and ${\cal A}$ means antisymmetrization with respect to all $z^{}_{1}$ to $z_{\ell N}^{}$. 

The wavefunction of the integer quantum Hall state $\nu_{\uparrow}=\nu_{\downarrow}=1$ is $\Psi^{(1,1)}([z], [w]) =\Phi[z] \Phi[w]$, where $[z]$ and $[w]$ are the coordinates of up and down spins. Next,  we note the identity (derived in Appendix)
\begin{eqnarray}
\Phi[z]\Phi[w]= C_{P}^{} {\cal A}^{}_{[z],[w]} \left( \prod_{N\geq i>j\geq 1}((R^{}_{P})_{i}-(R^{}_{P})_{j})^2 \right)  \hspace{0.2in} \label{CM1} \\
= C_{P}^{} {\cal A}^{}_{[z],[w]} \left( \Phi[R^{}_{P}]^2\right), \,\,\,\,\,\,\,\,   
(R^{}_{P})_{i}^{} \equiv (z^{}_{i}+w^{}_{Pi})/2,     \hspace{0.3in}   \label{RP}
 \end{eqnarray}
where $P$ is a permutation of the coordinates $[w]=(w_{1}^{}, w^{}_{2}, ..., w^{}_{N})$, and $C_{P}$ is a constant. Eq.(\ref{CM1}) shows that no matter how the opposite spins are grouped into pairs, their centers of mass automatically organize into a bosonic Laughlin state. 
In a projection sweep, the separation of an associating pair shrinks while its center of mass remains unchanged.  This fusing process can be modeled by the family
\begin{equation}
\Psi([z],[w]; t) = \sum_{P}{\cal A}^{}_{[z],[w]} \left[ \Phi[R^{}_{P}]^2 \prod_{i=1}^{N}f^{}_{t}((r^{}_{P})_{i}^{}) \right]
\label{family1}  \end{equation}
where $(r^{}_{P})^{}_{i} \equiv z^{}_{i}-w^{}_{Pi}$. 
The time $t=0$ and $t=1$ labels the beginning and the end of the sweep. $f_{t}(z_{i}-w_{Pi})$ is the wavefunction of the fermion pair in the grouping $P$ during the sweep.  
At $t=0$, $f^{}_{t=0}(r)=1$, so Eq.(\ref{family1}) reduced to the fermionic integer QH state $\Phi[z]\Psi[w]$. At $t=1$, $f_{t=1}$ is the wavefunction of a tightly bound molecule and can be treated as a delta-function on the scale of the spacing between molecules. The fermion coordinates $[z],[w]$ then reduce to the coordinates $[R^{}_{P}]$ of the bosonic molecules, and Eq.(\ref{family1}) becomes a sum of bosonic  wavefunctions $\Psi_{P}$ with coordinates  $\{ (R^{}_{P})_{i}^{} \}$, 
\begin{eqnarray}
\Psi([z],[w]; t=1)\rightarrow \sum_{P}\Psi_{P} \hspace{0.5in} \\
\Psi_{P} ={\cal S}\left[ \prod_{N\geq i>j\geq 1}((R^{}_{P})_{i}-(R^{}_{P})_{j})^2\right] 
\hspace{0.3in}
\end{eqnarray}
Note that different permutations corresponds to different ways to group opposite spins in to pairs, which leads to  a different set of centers of mass $\{ (R^{}_{P})_{i}^{} \}$. 
Because of the narrow width of $f_{t=1}$, the matrix element of any $n$-body operator of fermions between two different $P$ states in Eq.(\ref{family1}) will vanish unless $n$ is of order $N$. Thus, different states $\Psi_{P}$ belong to different Hilbert space. A measurement process will collapse the state in Eq.(\ref{family1}) into one of the bosonic Laughlin states. 

If the initial fermion state has more up than down spins, then the projection will give a boson-fermion QH mixture, 
\begin{equation}
\Psi_{BF}([R],[z]) = \prod_{i>j}( R_{i} - R_{j})^2 \prod_{i, a}(z_{a}-R_{i})\prod_{a>b}(z_{a}-z_{b}),
\label{BF} \end{equation}
where $R_{i}$ and $z_{a}$  are the coordinates of molecules and excess spins. If only a fraction of the fermions are fused into bosons and the rest remain in the integer QH state, the projected state is 
\begin{equation}
\Psi_{mix}([R],[z], [w])= \Phi[R]^2 \Phi[z]\Phi[w]\prod_{i, a}(z_{a}-R_{i})\prod_{i, b}(w_{b}-R_{i}).
\label{mix} \end{equation}
The density profiles of these states can be determined using standard plasma analog\cite{plasma}, which shows that the bosonic Laughlin state will sit at the center of the trap, with fermions surrounding it in a $\nu_{\uparrow}=1$  QH state. In the same way, one can create quasi-holes and their superposition in the bosonic Laughlin state by first engineering hole states and their superposition in the fermion $\nu_{\uparrow}=\nu_{\downarrow}=1$ state, and the sweep across the resonance at the projection rate. 
 
{\em Fusing of the integer fermion QH state $\nu_{\uparrow}=\nu_{\downarrow}=2$ :}
The wavefunction of the fermion integer QH state $\nu_{\uparrow}=\nu_{\downarrow}=2$ is, according to Eq.(\ref{ell}), 
\begin{equation}
\Psi^{(2,2)}([z],[w]) = {\cal A}_{z,w} \left( \prod_{k=0,1}\Phi[z^{(k)}]\Phi[w^{(k)}]
\prod_{i=1}^{N} z^{(1)\ast }_{i}w^{(1)\ast }_{i} \right). 
\label{22} \end{equation}
One can generalize Eq.(\ref{RP}) to two Landau levels so that the product of $\Phi$'s Eq.(\ref{22}) can be expressed in terms of centers of mass of the pairs. The generalization 
(derived in Appendix) is 
\begin{equation}
\prod_{k=0,1}\Phi[z^{(k)}]\Phi[w^{(k)}] = \overline{\cal A}\left(  
\Phi[ R_{P,Q}^{(0)}]^2 \Phi[ R_{P,Q}^{(1)}]^2 \right)
\label{CM2}\end{equation}
where $(R_{P,Q}^{(0)})_{i}\equiv \frac{1}{2}(z_{Pi}^{} + w^{}_{Qi})$,  $i=1,2, ..N$, and 
\begin{equation}
\Phi[ R_{P,Q}^{(0)}] =  \prod_{N\geq i>j \geq 1}
(z_{Pi}^{} + w^{}_{Qi} - z_{Pj}^{} - w^{}_{Qj})/2 \,\,\,\, . 
\label{R0} \end{equation}
$(R_{P,Q}^{(1)})_{i}$ and $\Phi[ R_{P,Q}^{(1)}]$ are similarly defined, obtained by replacing  $i$ and $j$ on the right hand side of $(R_{P,Q}^{(0)})_{i}$ and $\Phi[ R_{P,Q}^{(0)}]$  replaced by $i+N$ and $j+N$. Here, $P$ and $Q$ are arbitrary permutations of $2N$ coordinates of $z$ and  $w$ respectively. 
$\overline{\cal A}$ means antisymmetrization with respect to the sets $[z^{(0)}]$, $[z^{(1)}]$, $[w^{(0)}]$, and $[w^{(1)}]$. 
Eq.(\ref{CM2}) shows that for any grouping of the the up and down spins into pairs, the center of mass of these pairs will automatically organized into a 2-cluster (or bosonic Pfaffian) state under antisymmetrization. The distinction between Eq.(\ref{CM2}) and (\ref{CM1}) is that the up and down spins in Eq.(\ref{CM2})  can be from the first or the second Landau level (i.e. $k=0$ or $k=1$ in Eq.(\ref{ell})). These different ways to fuse the up and down spins into a boson is the origin of the non-abelian nature of the excitations of the Pfaffian state, which we shall discussed elsewhere. 

The projection family Eq.(\ref{family1}) is now generalized to 
\begin{eqnarray}
\Psi([z],[w]; t) = \sum_{P,Q}{\cal A}^{}_{[z],[w]} ( \prod_{N\geq i>j\geq 1}
((R^{}_{P,Q})_{i} - (R^{}_{P,Q})_{j})^2    \nonumber   \\
\prod_{N\geq i>j\geq 1} ((R^{}_{P,Q})_{i+N} - (R^{}_{P,Q})_{j+N})^2 
\hspace{1.0in}   
\nonumber \\
\prod_{i=1}^{N} z^{\ast}_{i+N}w^{\ast}_{i+N}   \prod_{i=1}^{2N} f^{}_{t}(z^{}_{Pi}-w^{}_{Qi})) 
 \hspace{1.0in}  
\label{family2} \end{eqnarray}
where $f_{t}$ is the same function as in Eq.(\ref{family1}). At $t=0$, 
Eq.(\ref{family2}) is the fermion integer QH state $\Psi^{(2,2)}([z],[w])$. As in the previous case
a measurement on Eq.(\ref{family2}) at $t=1$ will pick out a Pfaffian state of the form 
\begin{equation}
\Psi_{Pf}[R] = {\cal S}\left(\Phi[R^{(0)}]^2 \Phi[R^{(1)}]^2 t[R] \right)
\end{equation}
where $R^{(0)}_{i}\equiv R^{}_{i}$,  $R^{(1)}_{i}\equiv R^{}_{i+N}$, and 
${\cal S}$ is the symmetrization over the $2N$ bosons, and 
$t[R]$ is a polynomial of $R_{i}^{\ast}$ of degree $2N$ that comes from the $z^{\ast}$ products in Eq.(\ref{family2}).  
It has the same degree as a quasi-hole excitation $\prod_{i=1}^{2N}(R_{i}-a)$ (except for complex conjugation), and therefore has negligible effects on the underlying QH structure provided by the Laughlin factors, which has degree $(2N)^2$. 

It is straightforward to show that fermionic QH states with willing factor $\nu_{\uparrow}$ and $\nu_{\downarrow}$ 
will fuse into a bosonic state with filling factor  
\begin{equation}
\frac{1}{\nu_{B}^{}} = \frac{1}{\nu_{\uparrow}^{}} +  \frac{1}{\nu_{\downarrow}^{}}.
\label{nuB} \end{equation}
This is because a boson sees the flux of both up and down spins, while its number is the same as each spin population. The filling factor of the bosonic Laughlin and Pfaffian states are 1/2 and 1 respectively.

{\em Fusing fermionic QH state $\nu_{\uparrow}=1$ and $\nu_{\downarrow}=2$:}
The fermion wavefunction prior to the sweep is, (assuming $2N$ fermions of each spin),
\begin{eqnarray}
\Psi^{(1,2)}([z],[w]) = {\cal A}_{z} \left(\Phi[z^{(0)}]\Phi[z^{(1)}] \prod_{1\geq i, j \geq 1}(z^{(0)}_{i}- z^{(1)}_{j}) \right)  \nonumber \\
{\cal A}_{w} \left( \Phi[w^{(0)}]\Phi[w^{(1)}]\prod_{i=1}^{N}
w^{(1) \ast}_{i} \right)  \hspace{0.5in}
\label{1,2} \end{eqnarray}
Using Eq.(\ref{CM2}), and following the discussions of the previous section, one sees that the state  $\Psi^{(1,2)}([z],[w])$ (Eq.(\ref{1,2}) will fuse into 
a bosonic ``$(221)$"  state 
\begin{equation}
\Psi^{(221)}[R] = {\cal S}\left( \Phi[R^{(0)}]^2 \Phi[R^{(1)}]^2 \prod_{i,j}(R^{(0)}_{i}-R^{(1)}_{j})t[R]\right)
\label{B221}\end{equation}
where $t[R]$ is a polynomial of degree $N$, which has negligible effect on the underlying QH structure as discussed before. As shown in Eq.(\ref{nuB}), the (221) state has filling factor 
$\nu_{B}=2/3$. 

The nomenclature $(mmn)$  is taken from electron physics\cite{Halperin}\cite{bilayer}. 
Here, it means the symmetrized product 
${\cal S}\left[ \Phi[R^{(0)}]^{m}\Phi[R^{(1)}]^{m}  \prod_{i,j}(R^{(0)}_{i}-R^{(1)}_{j})^{n}\right]$.  In bilayer QH system,  $R^{(0)}$ and $R^{(1)}$  refer to electrons in the upper and lower layer respectively. However, there is no anti-symmetrization of electron coordinates between the layer due to lack of layer coherence in the limit of weak layer tunneling.  In Eq.(\ref{B221}), the symmetrization is necessary. 
From the standard flux 
counting\cite{bilayer} or the plasma analog\cite{plasma}, one finds the filling factor of the $(mmn)$ state is 
$\nu=2/(m+n)$, which is 2/3 in the case $\Psi^{(221)}$, in agreement with Eq.(\ref{nuB}). 

The quasi-holes of $\Psi^{(221)}$ are of the form  ( $\prod_{i=1}^{N}(R^{(0)}_{i}-a)\Psi^{(221)}$ or 
$\prod_{i=1}^{N}(R^{(1)}_{i}-a)\Psi^{(221)}$ ); as well as   $\prod_{i=1}^{2N}(R_{i}-a)\Psi^{(221)}$.
They have  fractional charge $1/(m+n)$ and $2/(m+n)$ respectively\cite{bilayer}. In the case of the (221) state, 
the first type of excitation (charge 1/3) can be created by inserting a hole 
 in  the $\ell=1$ or $\ell=1$ Landau level of the down spin fermions prior to the sweep,  
(i.e. $\prod_{i=1}^{N}(w^{(0)}_{i}-a)\Phi[w]$ or $\prod_{i=1}^{N}(w^{(1)}_{i}-a)\Phi[w]$).  
The second type (charge 2/3) can be created by inserting 
a hole in the lowest Landau level of the up
spin fermions ($\prod_{i=1}^{2N}(z_{i}-a)\Phi[z]$).  

{\em Further Discussions} : (i) Stability: Even though
the projection sweep is a non-equilibrium process, the resulting bosonic Laughlin state is stable because it is 
an eigenstate of bosons with short range repulsion. As for the bosonic Pfaffian, there is numerical evidence that it is close to the ground state for filling factor $\nu=1$\cite{Gunn}.

(ii) Generalization to higher integer QH states :   It is striaghtforward to generalize Eq.(\ref{CM2}) to  arbitrary number of Landau levels, (see also Appendix). This shows that the center of mass of the pairs in any grouping of opposite spins in the integer QH state $\nu_{\uparrow}=\nu_{\downarrow}=n$ will organized into a bosonic $n$-cluster QH state.

(iii) Projection sweep versus  ``sewing": By the ``sewing" of two quantum Hall states  $\Psi^{(\nu_{\uparrow})}[z]$ and $\Psi^{(\nu_{\downarrow})}[w]$, we mean taking the product of these two states and then let $w_{i}\rightarrow z_{i}$. This process is different from Eq.(\ref{family1}) and (\ref{family2}), since it will not keep the center of mass of the pairs unchanged in the fusing process. 
Although these two procedures give the same bosonic Laughlin state for the case $\nu_{\uparrow}=\nu_{\downarrow}=1$,   they give different results for high Landau levels, 
$\nu_{\uparrow}=\nu_{\downarrow}=n\geq 2$, as easily verified for a system 
of a few particles. 

(iv) Parton models: 
In the early studies of QH effects, a number of authors had suggested a patron construction
of QH states\cite{Jain,Wen}. Within such construction, an electron is viewed as a composite of three different types of fictitious fermions (called partons), each of which is in an integer QH state. The physical QH state  of the electrons is obtained by the fusing the QH states of the partons. Likewise, a bosonic QH state can be viewed as the fusion of the QH states of two different types of partons. Here, we show that there is indeed 
a physical process to approach the parton picture, at least in the bosonic case. In our process, the ``partons" are now no longer fictitious particles, but real fermions with spin up and down.  However, unlike the parton discriptions\cite{Jain,Wen} which keep the partons in the lowest or a few low lying Landau levels, the fermions in the projection sweep must include a  large number of Landau levels in order to form a bound state. On the BEC side of the resonance, only the center of mass of a fermion pair can remain in the low lying Landau levels. A fermion itself cannot.

{\em Appendix:} Proof of Eq.(\ref{Cauchy}), (\ref{CM1}) and (\ref{CM2}):   The right hand side of all these equations are polynomials of $[z]$ and $[w]$. Since any antisymmetric polynomial of $(z_{1},.., , z_{N})$ must contain a Vandermont determinant $\Phi[z]$, these polynomials must contain a Vandermont determinant of the coordinates to be antisymmetrized, and hence the results in Eq.(\ref{Cauchy}), (\ref{CM1}) and (\ref{CM2}), as well as the generalization of Eq.(\ref{CM2}) to arbitrary number of Landau levels, after comparing the degree of the polynomials on both sides of the equation. 

{\em Acknowledgement:} I thank Yuan-Ming Lu and Jainendra Jain for calling my attention to the patron models of QH state. This work is supported by the NSF Grant DMR-0907366, the MURI Grant FP054294-D, and the NASA Granton Fundamental Physics 1541824.


\begin{thebibliography}{99}
\bibitem{Cornell-LLL} V. Schweikhard, I. Coddington, P. Engels, V. P. Mogendorff, and E. A. Cornell
Phys. Rev. Lett. 92, 040404 (2004).

\bibitem{Cooper} N.R.  Cooper, Advances in Physics 57, 539 (2008)

\bibitem{Jin}  C. A. Regal, M. Greiner, and D. S. Jin, Phys. Rev. Lett. {\bf 92}, 040403 (2004)

\bibitem{RR} N. Read and E. Rezayi, Phys. Rev. B 59, 8084 (1999); E.H. Rezayi and N. Read, Phys. Rev. B 79, 075306 (2009).

\bibitem{Busch}   T Busch, B -G Englert, K Rzazewski and M Wilkens, Found. Phys. {\bf 28}, 549 (1998)

\bibitem{plasma} R. B. Laughlin, Chapt. 7, {\rm The Quantum Hall Effect}, edited by R.E. Prange and S. M. Girvin (Springer, New York, 1990). 

\bibitem{Halperin} B.I. Halperin, Helv. Phys. Acta {\b f 56}, 75 (1983). 

\bibitem{bilayer} S. M. Girvin and A. H. MacDonald, Chapter 5, pp.161-224, {\em Novel Quantum Liquids in Low-Dimensional Semiconductor Structures}, edited by Sankar Das Sar,a and Aron Pinczuk (
Wiley, New York, 1995). 

\bibitem{Gunn} N.R. Cooper, N.K.Wilkin, J.M. Gunn, Phys. Rev. Lett. {\bf 87}, 120405-1 (2001). 

\bibitem{Jain} J.K. Jain, Phys. Rev. B {\bf 40}, 8079, (1989). 

\bibitem{Wen} Xiao-Gang Wen and A. Zee, Phys. Rev. B {\bf 58}, 15717 (1998),  
 Rev. B {\bf 60}, 8827 (1999).  

\end{thebibliography}
\end{document}